\documentclass[conference]{IEEEtran}
\IEEEoverridecommandlockouts
\usepackage{makeidx}  
\usepackage{latexsym}
\usepackage{amsmath, amssymb, ascmac}
\usepackage{amsfonts}
\usepackage{float}
\usepackage{citesort}
\usepackage{mathrsfs}

\newtheorem{de}{Definition}
\newtheorem{thm}{Theorem}
  
\newtheorem{rem}{Remark}
\newtheorem{cor}{Corollary}
\newtheorem{prop}{Proposition}

\renewcommand{\QED}{\hfill $\Box$}
\newcommand{\R}{R\'enyi }

\renewcommand{\v}[1]{\mbox{\boldmath$ #1 $}}

\newcommand{\sv}[1]{{\mbox{\scriptsize \boldmath$ #1 $}}}

\newcommand{\ca}[1]{{{\mathcal #1}}}

\newcommand{\acr}[2]{R_\infty\left(#1|#2\right)}
\newcommand{\ar}[1]{R_\infty\left(#1\right)}

\ifCLASSINFOpdf
\else
\fi
\begin{document}
%
\title{Secret Sharing Schemes Based on Min-Entropies}

\author{\IEEEauthorblockN{Mitsugu Iwamoto}
\IEEEauthorblockA{
  Center for Frontier Science and Engineering, \\
The University of Electro-Communications, Japan\\ 
    {\tt mitsugu@uec.ac.jp}}
\and
\IEEEauthorblockN{Junji Shikata}
\IEEEauthorblockA{Graduate School of Environment and Information Sciences, \\
Yokohama National University, Japan \\
 {\tt shikata@ynu.ac.jp}}
}


%


\maketitle

\begin{abstract}
Fundamental results on secret sharing schemes (SSSs) are discussed in the setting where security and share size are measured by (conditional) min-entropies. \\
\quad 
We first formalize a unified framework of SSSs based on (conditional) \R entropies, which includes SSSs based on Shannon and min entropies etc.\ as special cases. By deriving the lower bound of share sizes in terms of \R entropies based on the technique introduced by Iwamoto--Shikata, we obtain the lower bounds of share sizes measured by min entropies as well as by Shannon entropies in a unified manner. \\
\quad 
As the main contributions of this paper, we show two existential results of non-perfect SSSs based on min-entropies under several important settings. We first show that there exists a non-perfect SSS for arbitrary {\em binary} secret information and arbitrary monotone access structure. In addition, for every integers $k$ and $n$ ($k \le n$), we prove that the ideal non-perfect $(k,n)$-threshold scheme exists even if the distribution of the secret is not uniformly distributed. 
\end{abstract}


%
\IEEEpeerreviewmaketitle

\section{Introduction}
A secret sharing scheme (SSS) \cite{S-cacm79,B-afips79} is one of the most fundamental primitives in cryptography. 
In  SSSs,  a secret information is encrypted into several information called {\em shares}, each of which has no information on the secret in the sense of information theoretic security. Each share is distributed to a participant, and the secret can be recovered by collecting the shares of specified set of participants called a {\em qualified set}. 

In a narrow sense, information theoretic security implies so-called {\em perfect security} \cite{S-BTJ49}. For instance, SSSs require that no information can be obtained by the non-qualified set of participants, called a {\em forbidden set}, even if they have unbounded computing power. Letting $S$ and $V_\ca{F}$ be the random variables corresponding to  the secret and the set of shares of a forbidden set $\ca{F}$, respectively, this requirement is mathematically formulated as $H(S|V_\ca{F}) = H(S)$ where $H(\cdot)$ and $H(\cdot|\cdot)$ are Shannon and conditional entropies, respectively. Namely, the random variables $S$ and $V_\ca{F}$ are statistically independent. 

On the other hand, another milder and optimistic scenario is studied to define the information theoretic security: Suppose that an adversary  can guess a plaintext  only once. Then, the best way to do this is guessing the plaintext  with the highest probability given the ciphertext. Merhav \cite{M-it03} studied the exponent of this kind of success probability in guessing for symmetric-key cryptography with {\em variable-length keys}\footnote{In this symmetric-key cryptography, each key depends on the ciphertext, and hence, its length can be varied depending on the ciphertext.} in {\em asymptotic setup}. Recently, Alimomeni and Safavi-Naini \cite{AS-icits12} and Dodis \cite{D-icits12}  revisited the same scenario for the symmetric-key cryptography with {\em fixed-length keys} such as Vernam cipher \cite{Vernam-jaiee26} in {\em non-asymptotic setup}. The security criteria in  \cite{AS-icits12,D-icits12}  are based on {\em min-entropy} and its conditional version since the min-entropy is defined by negative logarithm of the highest probability in a probability distribution. Under such security criteria, the lower bounds of key length are discussed in \cite{AS-icits12,D-icits12}. 

In this paper, we are interested in SSSs by using (conditional) min-entropies, and clarify their fundamental results. Particularly, we investigate the lower bounds of share sizes, and the constructions of SSSs based on min-entropies. 

In order to  derive the lower bounds of share sizes in terms of min-entropies, we take the similar strategy developed recently by Iwamoto and Shikata \cite{IS-icits13}. Namely, we first formalize a unified framework of SSSs based on (conditional) \R entropies, which includes SSSs based on Shannon and min entropies etc., as special cases.  Then, by deriving the lower bound of share sizes in terms of \R entropies, we can obtain the lower bounds of share sizes measured by Shannon and min entropies in a unified manner. 

Then, we show two existential results on SSSs based on min-entropies. Noticing that SSSs satisfying perfect security also satisfy the security criteria based on min-entropies, we are particularly interested in so-called {\em non-perfect} SSSs \cite{BM-crypto85,HY-ieice,KOSOT-ecrypt93} while they are secure in the sense of min-entropies. As a result, we clarify the following two fundamental facts: 

The first result is on the existence of SSSs with {\em general access structures}.  In  SSSs satisfying perfect security, Ito, Saito, and Nishizeki \cite{ISN-jc93} proved the well known result that SSSs can be constructed if and only if the access structure satisfies a certain property called {\em monotone}. Combining this result with the one by  Blundo, De Santis, and Vaccaro \cite{BSV-ipl98}, this existential result can be extended to the case of arbitrary distribution of secret information. Inspired by  these results,  we will clarify that we can always construct non-perfect SSS based on min-entropies for arbitrary monotone access structures and arbitrary {\em binary} probability distribution of the secret. 

The second result is on the optimality of SSSs based on min-entropies. In SSSs with perfect security, SSS is called {\em ideal} if $H(S)=H(V_i)$ for all $i=1,2,\ldots,n$. Note that the ideal SSS only exists only in the case where $S$ is uniformly distributed \cite{BSV-ipl98}.  In this case, $\ar{S}=\ar{V_i}$ obviously holds since Shannon and min entropies coincide for uniform distributions. Hence, we are interested in the existence of ideal non-perfect SSSs based on min-entropies for {\em non-uniform}  probability distribution of the secret. Surprisingly, we will prove that there actually exists such a non-uniform probability distribution of the secret realizing the ideal non-perfect $(k,n)$-threshold  SSS based on min-entropies. 

The remaining part of this paper is organized as follows: In Section \ref{model_SS}, we provide a formal model and security definition of SSSs based on \R entropies for general access structures. Under such a  model and security definitions, we derive the lower bound of share sizes measured by average conditional min-entropies in Section \ref{sec:LB}, by proving the extended lower bound based on \R entropies. Sections \ref{sec:AS} and \ref{sec:ideal} are devoted to construct SSSs based on min-entropies. In Section \ref{sec:AS}, we show the existence of non-perfect SSSs based on min-entropies for {\em arbitrary} distribution on {\em binary} secret and for arbitrary monotone access structures. Then,  it is clarified in Section \ref{sec:ideal} that an ideal non-perfect $(k,n)$-threshold SSS based on min-entropies exists for non-uniform distribution of the secret. 
Technical proofs of Theorems \ref{thm:n-out-of-n-SSS} and  \ref{thm:nontrivaial_ideal_SSS} are provided in Appendices \ref{sec:n-out-of-n-SSS} and \ref{app:nontrivaial_ideal_SSS}, respectively. 

\section{Model and Security Definition}\label{model_SS}   
Let $[n]:=\{ 1,2,\ldots, n \}$ be a finite set of IDs of $n$ users. For every $i\in [n]$, let ${\cal V}_i$ be a finite set of possible shares of the user $i$, and denote by $P_{V_i}$ its associated probability distribution on ${\cal V}_i$. Similarly, let $\ca{S}$ be a finite set of secret information and $P_{S}$ be its associated probability distribution.  In the following, for any subset $\ca{U}:=\{i_1,i_2, \ldots,i_u \}  \subset [n]$, we use the notation $v_\ca{U}:=\{v_{i_1}, v_{i_2},\ldots, v_{i_u}\}$ and $V_\ca{U}:=\{V_{i_1}, V_{i_2},\ldots, V_{i_u}\}$.  For a finite set $\ca{X}$, let ${\mathscr P}(\ca{X})$ be the family of probability distributions defined on $\ca{X}$. 

\subsection{Access Structures} 
In SSSs, we normally assume that each set of shares is classified into either a {\em qualified set} or a {\em forbidden set}. 
A {\em qualified set} is the set of shares that can recover the secret. On the other hand, the secret must be kept secret against any collusion of members of a forbidden set in the sense of information theoretic security, the meaning of which will be formally specified later. 
Let $\mathscr{Q}\subset 2^{\ca{[n]}}$ and $\mathscr{F}\subset 2^{\ca{[n]}}$ be families of qualified and forbidden sets, respectively. Then we call $\varGamma:=(\mathscr{Q}, \mathscr{F})$ an {\em access structure}. In particular, the access structure is called {\em $(k,n)$-threshold access structure} if it satisfies that 
$
\mathscr{Q} := \{\ca{Q}:|\ca{Q}|\ge k\} 
$ and 
$\mathscr{F} := \{\ca{F}:|\ca{F}|\le k-1\}$. In this paper, we assume that the access structure is a partition of $2^{[n]}$, namely, ${\mathscr Q} \cup {\mathscr F} = 2^{[n]}$ and ${\mathscr Q} \cap {\mathscr F} = \emptyset$.

An access structure is said to be {\em monotone} if it satisfies that 
for all $\ca{Q}\in\mathscr{Q}$, every $\ca{Q}' \supset \ca{Q}$ satisfies $\ca{Q}'\in\mathscr{Q}$ and;
for all $\ca{F}\in\mathscr{F}$, every $\ca{F}' \subset \ca{F}$ satisfies $\ca{F}'\in\mathscr{F}$. In addition, we define the {\em maximal forbidden sets} as $\mathscr{F}^+ := \{\ca{F} \in \mathscr{F}\mid \ca{F} \cup \{i\}\in \mathscr{Q}, \mbox{ for all } i\in [n]\backslash\ca{F}\}$.  Clearly, the monotone property is necessary condition for the existence of secret sharing schemes. Furthermore, it is proved that this property is actually sufficient for the existence of SSSs \cite{ISN-jc93}.

\subsection{Secret Sharing Schemes for General Access Structures}
Let $\varPi=([P_{S}], \varPi_{share}, \varPi_{comb})$ be a {\it secret sharing scheme for an access structure $\varGamma$}, as defined below:  
\begin{itemize}
\item $[P_{S}]$ is a sampling algorithm for secret information, and it outputs a secret $s\in \ca{S}$ according to a probability distribution $P_{S}$; 
\item $\varPi_{share}$ is a randomized algorithm for generating shares for all users, and it is executed by a honest entity called {\it dealer}. It takes a secret $s\in \ca{S}$ on input and outputs $(v_1,v_2,\ldots,v_n)\in \prod_{i=1}^{n} \ca{V}_i$; and 
\item  $ \varPi_{comb}$ is an algorithm for recovering a secret. It takes a set of shares  $v_\ca{Q}$, $\ca{Q} \in \mathscr{Q}$, on input and outputs a secret $s\in \ca{S}$.
\end{itemize}

In this paper, we assume that $\varPi$ meets {\it perfect correctness}: for any possible secret $s\in \ca{S}$, and for all possible shares $(v_1,v_2,\ldots,v_n)\gets \varPi_{share}(s)$,  it holds that $\varPi_{comb}(v_\ca{Q})=s$ for any subset $\ca{Q}\in\mathscr{Q}$.

\subsection{Information Measures and Security Criteria}

In information theoretic cryptography, information measures play a significant role since it is used to define the security as well as to measure the share sizes. In this paper, we are interested in \R entropy and its conditional one since \R entropies include several useful entropy measures as special cases. 

For a non-negative real number $\alpha$ and a random variable $X$ taking its values on a finite set $\ca{X}$, \R entropy  of order $\alpha$ with respect to $X$ is defined as \cite{R-BSMS61}
\begin{align}
R_\alpha(X):=\frac{1}{1-\alpha}\log \sum_{x\in\ca{X}} P_X(x)^\alpha. 
\end{align}
It is well known several entropy measures are special cases of \R entropy. For instance, Shannon entropy $H(X):=-\sum_{x\in \ca{X}}P_X(x)\log P_X(x)$ and min-entropy $R_\infty(X):=
-\log\max_{x\in \ca{X}} P_X(x)$ are derived as special cases of  $R_\alpha(X)$ by letting $\alpha \rightarrow 1$ and $\alpha \rightarrow \infty$, respectively.  

In addition, for a non-negative real number $\alpha$ and random variables $X$ and $Y$ taking values on  finite sets $\ca{X}$ and $\ca{Y}$, respectively, conditional \R entropy  of order $\alpha$ with respect to $X$ given $Y$ is defined as \cite{A-CMSJB75}
\medskip

$R_\alpha(X|Y)$
\vspace{-4mm}
\begin{align}
\label{eq:H-Renyi}
:=\frac{\alpha}{1-\alpha}\log \sum_{y\in \ca{Y}}P_Y(y)\left\{\sum_{x\in\ca{X}} P_{X|Y}(x|y)^\alpha\right\}^{1/\alpha}. 
\end{align}


Note that there are many definitions of conditional \R entropies \cite{TAML-it12,IS-icits13}. There are two reasons  why we choose the definition \eqref{eq:H-Renyi} of conditional \R entropy. The first reason is that it is connected to the cryptographically important conditional min-entropy $\acr{\cdot}{\cdot}$ (e.g., see \cite{DKRS-08}) defined as
\begin{align}
\acr{X}{Y} := -\log \sum_{y \in \ca{Y}} P_Y(y) \max_{x\in\ca{X}} P_{X|Y}(x|y)
\end{align}
which plays a crucial role in this paper. We note that the relation $\acr{X}{Y}=\lim_{\alpha \rightarrow \infty} R_\alpha(X|Y)$ holds \cite{IS-icits13}.  

The second reason is that it satisfies the very useful properties similar to Shannon entropy as shown below: 
\begin{prop}[\cite{A-CMSJB75,A-it96,IS-icits13}]\label{prop:basic-properties}
Let $X$, $Y$, and $Z$ be random variables taking values on finite sets $\ca{X}$, $\ca{Y}$, and $\ca{Z}$, respectively. 
Then, for arbitrary $\alpha \in [0,\infty]$ we have: 
\begin{itemize}
\item[(A)] {\em Conditioned Monotonicity:} 
\begin{itemize}
\item[(i)] $R_{\alpha}(X|Z)\le R_{\alpha}(XY|Z)$,  and
\item[(ii)] $R_{\alpha}(X|Z)= R_{\alpha}(XY|Z)$  holds if and only if $Y=f(X,Z)$ for some (deterministic) mapping $f$.
\end{itemize}
\item[(B)] {\em Conditioning Reduces Entropy:} $R_\alpha(X)\ge R_\alpha(X|Y)$ \\ 
where the equality holds if $X$ and $Y$ are statistically independent. 
\end{itemize}
\end{prop}

Note that the properties (A)--(i) and (B) are proved in \cite{A-it96} and \cite{A-CMSJB75}, respectively. The property (A)--(ii) is explicitly pointed out in \cite{IS-icits13}. 

Based on the above foundations of (conditional) \R entropies, we give security formalization for a secret sharing scheme for an access structure $\varGamma$ as follows. 

\begin{de}[Security based on \R entropies]\label{de:security} \label{de:SS_security_Renyi} Let $\varPi$ be a secret sharing scheme for an access structure $\varGamma$. 
Then,  $\varPi$ is said to meet {\it $\epsilon$-security with respect to $R_{\alpha}(\cdot|\cdot)$}, if for any forbidden set $\ca{F} \in \mathscr{F}$ it satisfies 
\begin{align}\label{eq:SSS_sec}
R_{\alpha}(S)-R_{\alpha}(S|V_{\ca{F}})\le \epsilon. 
\end{align} 
In particular, $\varPi$  is said to meet {\it perfect security with respect to $R_{\alpha}(\cdot|\cdot)$} if $\epsilon=0$ above. 

For simplicity we abbreviate $\epsilon$-security with respect to $R_{\alpha}(\cdot|\cdot)$ as $(R_\alpha(\cdot|\cdot),\epsilon)$-security, and perfect security with respect to $R_{\alpha}(\cdot|\cdot)$ as $R_\alpha(\cdot|\cdot)$-security. 
\end{de}  

Note that Definition \ref{de:security} includes the security of SSSs based on several types of entropies as special cases. 
In particular, we are interested in the case of $\alpha \rightarrow 1$ and $\alpha\rightarrow \infty$ which define $\epsilon$-security with respect to $H(\cdot|\cdot)$ and $R_\infty(\cdot|\cdot)$, i.e., the security based on Shannon and min entropies, respectively. 

Finally, we note that, if a SSS is $H(\cdot|\cdot)$-secure, i.e., perfectly secure,  it is $R_\infty(\cdot|\cdot)$-secure as well. Hence, in this paper, we are mainly interested in the SSS satisfying $R_\infty(\cdot|\cdot)$-security but not satisfying $H(\cdot|\cdot)$-security. Such a SSS is called {\em non-perfect} SSS \cite{KOSOT-ecrypt93}, which is formally defined as follows:
\begin{de}[\cite{KOSOT-ecrypt93}]\label{de:non-perfect}
If there exists a set of $\ca{F} \subset [n]$  satisfying  
$H_{\alpha}(S) > H_{\alpha}(S|V_{\ca{F}})$, it is called a {\em non-perfect} SSS. 
\end{de}

\section{Unified Proofs for Lower Bounds \\ of Share Sizes in SSSs}  \label{sec:LB}

We begin with this section by reviewing the following well-known tight lower bound of SSSs for arbitrary access structure $\varGamma$ meeting $(H(\cdot|\cdot),\epsilon)$-secure SSSs. 
\begin{prop}[\cite{KGH-it83,CSGV-jc93}] \label{prop:known_SSbound}
Let $\varPi$ be an $H(\cdot|\cdot)$-secure secret sharing scheme for an access structure $\varGamma$. 
Then, for every $i\in [n]$, it holds that   
\begin{align}\label{eq:known_SSbound} 
H(V_i)\ge H(S) \mbox{ and } |\ca{V}_i| \ge |\ca{S}|.
\end{align}
\end{prop}

Note that \eqref{eq:known_SSbound} is proved for threshold access structures and general access structures in \cite{KGH-it83} and \cite{CSGV-jc93}, respectively. From Proposition \ref{prop:known_SSbound}, we define that the secret sharing scheme $\varPi$ is called {\em ideal} if $\varPi$ satisfies \eqref{eq:known_SSbound} {\em with equalities}. 

As is shown in Proposition \ref{prop:known_SSbound}, we note that the SSSs based on Shannon entropy are well studied. On the other hand, there is {\em no} study on a SSS based on min-entropy, which is the main topic of this paper.  Noticing the fact that \R entropy is a generalization of Shannon, min, and several kinds of entropies, it is fruitful to derive the lower bounds of share sizes of 
$(R_{\alpha}(\cdot|\cdot),\epsilon)$-secure SSSs, i.e., in terms of \R entropies, in a comprehensive way. 
Then, we can directly prove that the lower bounds of share sizes of 
$(R_{\infty}(\cdot|\cdot),\epsilon)$-secure SSSs as a corollary. 
The following theorem can be considered as an extension of impossibility result with respect to key sizes measured by \R entropy discussed in \cite{IS-icits13} in symmetric-key cryptography to the impossibility result with respect to SSSs. 

\begin{thm} \label{thm:SS_bound1} 
Let $\varPi$ be a secret sharing scheme for an access structure $\varGamma$ which meets 
$(R_{\alpha}(\cdot|\cdot),\epsilon)$-security. 
Then, it holds that, for arbitrary $\alpha \in [0, \infty]$, 
\begin{align}\label{eq:SSS_lowerbound}
R_{\alpha}(V_i)\ge R_{\alpha}(S)-\epsilon 
\end{align}
for every $i\in [n]$. 
\end{thm}

{\it Proof.} 
For any $i\in [n]$, there exists a for bidden set $\ca{F}\in \mathscr{F}$ such that $i\not\in \ca{F}$ and $\ca{F} \cup \{i\}\in \mathscr{Q}$. Then, we have 
\begin{align}
R_{\alpha}(S)&\le R_{\alpha}(S|V_{\ca{F}})+\epsilon  \nonumber \stackrel{\mbox{\scriptsize (a)}}\le R_{\alpha}(SV_{i}|V_{\ca{F}})+\epsilon \\
&\stackrel{\mbox{\scriptsize (b)}}= R_{\alpha}(V_i| V_{\ca{F}})+\epsilon \stackrel{\mbox{\scriptsize (c)}}\le R_{\alpha}(V_i)+\epsilon, \label{SSbound1-3}
\end{align}
where (a) follows from Proposition \ref{prop:basic-properties} (A)--(i), (b) follows from Proposition \ref{prop:basic-properties} (A)--(ii)  since $\varPi$ meets perfect correctness, and (c) follows from Proposition \ref{prop:basic-properties} (B). \hfill \QED

In the case of $R_\alpha(\cdot|\cdot)$-security, i.e., $\epsilon =0$,  the following corollary  obviously holds. 
\begin{cor} \label{cor:SS_renyi1}
For arbitrarily fixed $\alpha \in [0,\infty]$, let $\varPi$ be a secret sharing scheme for an access structure $\varGamma$ meeting $R_\alpha(\cdot|\cdot)$-security. Then, it holds that   
$R_{\alpha}(V_i) \ge R_{\alpha}(S)$  
for every $i\in [n]$.
\end{cor} 

Hence, we can immediately obtain the following corollary with respect to the lower bound of share sizes measured by {\em  min-entropy} under {\em $\acr{\cdot}{\cdot}$-security} by letting $\alpha \rightarrow \infty$ in Corollary  \ref{cor:SS_renyi1}. 

\begin{cor} \label{cor:SS_renyi2}
Let $\varPi$ be a secret sharing scheme for an access structure $\varGamma$ meeting $\acr{\cdot}{\cdot}$-security. Then,  it holds that   
$R_{\infty}(V_i) \ge R_{\infty}(S)$  
for every $i\in [n]$.
\end{cor} 

Moreover, noting that $R_\alpha(S|V_\ca{F}) = R_\alpha(S)$ holds for {\em arbitrary} $\alpha  \in [0,\infty]$ {\em if} the random variables $S$ and $V_\ca{F}$ are statistically independent\footnote{However, as  will be shown in Sections \ref{sec:AS} and \ref{sec:ideal}, we note the converse of this implication is not always true.}, i.e., $H(S|V_\ca{F}) = H(S)$. Hence, we can prove that the lower bounds of share sizes of SSSs with $H(\cdot|\cdot)$-security. 

\begin{cor} \label{cor:SS_min_perfect1}
Let $\varPi$ be a secret sharing scheme for an access structure $\varGamma$ meeting $H(\cdot|\cdot)$-security. Then, it holds for arbitrary $\alpha  \in [0,\infty]$ that 
$R_{\alpha}(V_i) \ge R_{\alpha}(S)$, $i\in [n]$.
\end{cor} 

Hence, Proposition \ref{prop:known_SSbound} is immediately obtained from Corollary \ref{cor:SS_min_perfect1} as a special cases of Corollary \ref{cor:SS_min_perfect1} by taking the limits $\alpha \rightarrow 1$ and $\alpha \rightarrow 0$. 

In a similar manner, we can also obtain the following corollary with respect to the lower bound of share sizes measured by {\em  min-entropy} under traditional {\em $H(\cdot|\cdot)$-security}  by letting $\alpha \rightarrow \infty$ in Corollary  \ref{cor:SS_min_perfect1}. 

\begin{cor} \label{cor:SS_min_perfect2}
Let $\varPi$ be a secret sharing scheme for an access structure $\varGamma$ meeting $H(\cdot|\cdot)$-security. Then, it holds that   
$R_{\infty}(V_i) \ge R_{\infty}(S)$  
for every $i\in [n]$.
\end{cor} 

\begin{rem}
Recently, Alimomeni and Safavi-Naini \cite{AS-icits12} proved that the key size in symmetric-key encryption must be equal to or larger than the message size  if these sizes  are measured by {\em min-entropy} and the security criteria is based on $R_\infty(\cdot|\cdot)$. Similarly, Dodis \cite{D-icits12} proved that the  key size in symmetric-key encryption\footnote{Note that the probabilistic encryption is considered in the result by Dodis while only deterministic encryption is discussed in the result by Alimomeni and Safavi-Naini. In SSSs, we do not care such a difference since $\Pi_{share}$ (i.e., the encryption function of SSSs) is randomized in nature. } must be equal to or larger than the message size  if these sizes are measured by {\em min-entropy} and the security criteria is based on $H(\cdot|\cdot)$.  Hence, Corollaries \ref{cor:SS_renyi2} and \ref{cor:SS_min_perfect2} can be considered as SSS versions of \cite{AS-icits12} and \cite{D-icits12}, respectively. 
\end{rem}

\begin{rem}
All discussions in Sections \ref{model_SS} and \ref{sec:LB} are valid if we replace $R_\alpha(\cdot|\cdot)$ with the conditional \R entropy proposed by  Hayashi \cite{H-it11}. In this case, the only difference is the resulting conditional min-entropy defined as $R^{\sf wst}_\infty(X|Y):=-\log \max_{(x,y)\in\ca{X} \times \ca{Y}} P_{X|Y}(x|y)$. However, even if we give the security definition based on  Hayashi's conditional \R entropy, we can also prove Theorem \ref{thm:SS_bound1} in the same way. 
\end{rem}

\section{Existence of Non-Perfect SSSs Based on Min-entropies for General Access Structures}\label{sec:AS}
Hereafter, we are concerned with the existence of SSSs satisfying $R_\infty(\cdot|\cdot)$-security. 
Recalling the discussion on Definition \ref{de:non-perfect}, we are concerned with a $\acr{\cdot}{\cdot}$-secure {\em non-perfect} SSS, since, if a SSS is perfectly secure, it is $R_\infty(\cdot|\cdot)$-secure as well. 
Note that non-perfect SSSs are meaningless if the secret if deterministic since $H(S)=0$ in such a  case. Hence, we assume that the secret is not deterministic in the following discussion. 

In this section, we address  the existence of non-perfect SSS satisfying $\acr{\cdot}{\cdot}$-security for arbitrary monotone access structure $\varGamma$. 

\subsection{Existence of Non-perfect SSS Based on Min-entropies}
Combining the results \cite{ISN-jc93} and \cite{BSV-ipl98}, there exist SSSs satisfying $H(\cdot|\cdot)$-security for arbitrary monotone access structure and for {\em arbitrary} probability distribution of secret information. 
However, it is not known whether this fact is still valid or not for {\em non-perfect} SSSs satisfying  $R_\infty(\cdot|\cdot)$-security. 
We obtain a positive result for such a question, summarized as the following theorem:

\begin{thm}\label{thm:existence}
For  an arbitrary {\em binary} and {\em non-uniform} probability distribution $P_S(\cdot) \in \mathscr{P}(\{0,1\})$ of secret information, there exists a non-perfect SSS with $\acr{\cdot}{\cdot}$-security with an arbitrary monotone access structure $\varGamma=(\mathscr{Q},\mathscr{F})$. 
\end{thm}

{\em Proof.} Let $P_S(\cdot)\in \mathscr{P}(\{0,1\})$ be an arbitrarily fixed non-uniform probability distribution of the binary secret. 
For a given monotone access structure $\varGamma=(\mathscr{Q},\mathscr{F})$, let $\mathscr{F}^+:=\{\ca{F}_1, \ca{F}_2, \ldots, \ca{F}_m\}$ where $m:=|\mathscr{F}^+|$. 

Suppose that we can generate a set of shares denoted by $\{w_1,w_2,\ldots,w_m\}$ of non-perfect $(m,m)$-threshold  SSS for the secret $s\in\{0,1\}$ satisfying $\acr{\cdot}{\cdot}$-security. The construction of such $(m,m)$-threshold SSS based on min-entropies will be provided in Construction $\varPi_1$. For $j \in [m]$, let $W_j$ be the random variable taking its values on a finite set $\ca{W}_j$, which corresponds to $w_j$. 

Consider a {\em cumulative map} \cite{ISN-jc93} $\varphi^\mathscr{F}: [n] \rightarrow 2^{[m]}$ given by 
$\varphi^\mathscr{F}(i) :=\{j\mid i \not \in \ca{F}_j \in \mathscr{F}^+\}$ for $i\in[n]$, and define $\varphi^\mathscr{F}(\ca{U}) :=\bigcup_{i\in\ca{U}}\varphi^\mathscr{F}(i)$ for $\ca{U} \subset [n]$.  
Then, it is  proved in \cite{ISN-jc93} that 
\begin{align}
\label{eq:prop_cmp1}
\left|\varphi^\mathscr{F}(\ca{U})\right| &\ge m, &\mbox{if } &\ca{U} \in {\mathscr Q}\\
\label{eq:prop_cmp2}
\left|\varphi^\mathscr{F}(\ca{U})\right| &\le m-1, &\mbox{if } &\ca{U} \in {\mathscr F}.
\end{align}

Now, we assume that each share $v_i$ for the SSS with the access structure $\varGamma$ consists of a set of $w_j$. Specifically, let $v_i := \{w_j\mid j \in \varphi^\mathscr{F}(i)\}$. Then, the secret $s$ can be recovered from a qualified set $\ca{Q}\subset[n]$ due to  \eqref{eq:prop_cmp1}.  On the other hand, we have 
\begin{align}\label{eq:sec_proof}
\acr{S}{V_\ca{F}} = \acr{S}{W_{\varphi^\mathscr{F}(\ca{F})}} = R_\infty(S)
\end{align}
holds for arbitrary $\ca{F} \in \mathscr{F}$ where the first equality holds from the definition of $v_i$, and the second equality is due to \eqref{eq:prop_cmp2} and $\acr{\cdot}{\cdot}$-security for the non-perfect $(m,m)$-threshold SSS. \QED

\begin{rem}
Similar argument also holds by combining monotone circuit construction \cite{BL-ecrypt88} with $(n,n)$-threshold non-perfect SSS with $\acr{\cdot}{\cdot}$-security, which is omitted here. 
\end{rem}

Hence, the remaining to prove Theorem \ref{thm:existence} is the construction of a non-perfect $(n,n)$-threshold SSS satisfying $\acr{\cdot}{\cdot}$-security for {\em arbitrary} non-uniform probability distribution $P_S(\cdot)\in\mathscr{P}(\{0,1\})$. 

\subsection{Construction of Non-perfect $(n,n)$-threshold SSS Based on Min-entropies for Arbitrary Binary Secret Information}\label{sec:n-out-of-n}
\noindent
{\em Construction $\varPi_1$:} Let $S$, and $V_1,V_2,\ldots,V_n$ be binary random variables. Assume that $S$ and $V_1,V_2,\ldots,V_{n-1}$ are statistically independent and they satisfy
$P_S(0)=P_{V_1}(0)=\cdots=P_{V_{n-1}}(0)=p$, for $1/2 < p < 1$. 
Then, we generate $V_n$ by $V_n := S\oplus V_1 \oplus V_2 \oplus \cdots \oplus V_{n-1}$ where $\oplus$ denotes the exclusive OR operation. 

\begin{thm}\label{thm:n-out-of-n-SSS}
The construction $\varPi_1$ realizes 
\begin{align}
\ar{S} &= \ar{V_i} = - \log p \ 
\label{eq:ideality}
\mbox{ for }~ i\in[n-1],\\
\label{eq:non-ideality}
\ar{V_n} &= -\log\{p^2+(1-p)^2\},
\end{align}
and
\begin{align}\label{eq:security_pi1}
\hspace{-2mm}\acr{S}{V_\ca{F}} &= -\log p \ \mbox{ for } \ca{F}\subset [n]~\mbox{s.t.\ } |\ca{F}|=n-1.
\end{align}
Hence, the construction $\varPi_1$ is a non-perfect $(n,n)$-threshold SSS which meets $R_{\infty}(\cdot | \cdot)$-security. 
\end{thm}

{\em Proof.} 
See Appendix \ref{sec:n-out-of-n-SSS}.  \QED

\begin{rem}\label{rem:exception}
The above construction $\varPi_1$ works in the cases of $p=1/2, 1$. In both cases, the random variables $S$ and $V_{\ca{F}}$ in the construction $\varPi_1$ are statistically independent, and hence, they result in $(n, n)$-threshold SSSs satisfying $H(\cdot|\cdot)$-security. 

On the other hand, if the random variables $S$ and $V_{\ca{F}}$ in the construction $\varPi_1$ are {\em not} statistically independent if $n\in \ca{F}$ since  $p \neq 1/2$, $1$. Therefore,  $\varPi_1$ is a non-perfect $(n, n)$-threshold SSS while it satisfies $R(\cdot|\cdot)$-security. 
\end{rem}

\begin{rem}
As we will see in Section \ref{sec:ideal}, there exists a specific probability distribution of the secret that realizes a non-perfect SSS satisfying $\acr{\cdot}{\cdot}$-security for an arbitrary monotone access structure $\varGamma=(\mathscr{Q},\mathscr{F})$ even if the set $\ca{S}$ is non-binary. However, it is an open problem to construct a non-perfect SSS for $\varGamma$ satisfying $\acr{\cdot}{\cdot}$-security  for arbitrary probability distribution $P_S$ on $\cal S$ if $|\ca{S}| \ge 3$. 
\end{rem}

Now, we analyze the efficiency of the protocol $\varPi_1$. Note that in the case of ordinary SSSs satisfying $H(\cdot|\cdot)$-security, the SSS satisfying $H(V_i)=H(S)$ for all $i \in [n]$ is called {\em ideal} since such a SSS is considered to be optimal in terms of share sizes due to \eqref{eq:known_SSbound} in Proposition \ref{prop:known_SSbound}. 

In our problem setting, we can use Corollary \ref{cor:SS_renyi2} to define the {\em ideal} SSS  for $R_\infty(\cdot|\cdot)$-security. 
\begin{de}[Ideal SSS based on Min-entropies]
A SSS $\varPi$ meeting $\acr{\cdot}{\cdot}$-security is called {\em ideal} 
if $\varPi$ satisfies $R_\infty(V_i) = R_\infty(S)$ for arbitrary $i \in [n]$. 
\end{de}

Based on the above definition, the share sizes in $\varPi_1$ are ``almost'' ideal in the sense that $R_\infty(V_i)=R_\infty(S)$ for $i\in[n-1]$, but $R_\infty(V_n) > R_\infty(S)$. 
In order to obtain (fully) ideal SSS which meets $R_{\infty}(\cdot | \cdot)$-security, the parameter $p$ must satisfy 
$
-\log(p^2+(1-p)^2)= -\log p
$ 
because of $\ar{S}=\ar{V_n}$, and we have $p=1/2,1$. 
However, each of $p=1/2,1$ makes $\varPi_1$ to be $H(\cdot|\cdot)$-secure as pointed out in Remark \ref{rem:exception} (and hence, we assume $p \neq 1/2,1$ in $\varPi_1$). Namely, $\varPi_1$  cannot realize {\em ideal non-prefect} $\acr{\cdot}{\cdot}$-secure SSSs while it is applicable to arbitrary probability distribution of binary secret. 

Summarizing, although the protocol $\varPi_1$ is applicable to {\em arbitrary} binary probability distribution of the secret, it has the following problems; $\varPi_1$ is designed for $(n,n)$-threshold schemes with binary secrets, and; $\varPi_1$ cannot realize {\em ideal non-prefect} SSSs satisfying $\acr{\cdot}{\cdot}$-security. 

\section{Existence of Ideal Non-perfect $(k,n)$-threshold SSS Based on Min-entropies}\label{sec:ideal}



From the discussion at the end of the last section, we show that there exist an ideal  non-perfect $(k,n)$-threshold SSSs satisfying $\acr{\cdot}{\cdot}$-security with specific {\em non-uniform} probability distributions of a secret $S$ over arbitrary finite field $\ca{S}$. This result implies the essential difference between $\acr{\cdot}{\cdot}$- and $H(\cdot|\cdot)$-security since  it is proved in \cite[Theorem 7]{BSV-ipl98} that ideal $H(\cdot|\cdot)$-secure SSS is realized only when $S$ is uniform.

\noindent
{\em Construction $\varPi_2$:} 
For a finite field $\mathbb{F}_t$ with a prime power $t$,  generate a set 
\begin{align}
\nonumber
\hspace*{-.5mm}
{\sf DT}_{(k,n)}:=\Big\{&(s,v_1,v_2,\ldots,v_n) \mid 
v_i = s + \sum_{\ell=1}^{k-1} i^\ell r_\ell , \\
\label{eq:relation}
\hspace*{-.5mm}
&(s,r_1,r_2,\ldots,r_{k-1})\in \left(\mathbb{F}_t\right)^k\Big\}\subset \left(\mathbb{F}_t\right)^{n+1}
\end{align}
called {\em distribution table} \cite{Stinson-05}, where we assume that each $i\in[n]$ in \eqref{eq:relation} is appropriately encoded  so as to be regarded as $[n] \subset \mathbb{F}_t$. 
Let $S$, and $V_1,V_2,\ldots,V_n$ be  random variables with joint probability  given by 
\begin{align}
\nonumber
&P_{SV_1V_2\cdots V_n}(s,v_1,v_2,\ldots,v_n)\\
\label{eq:(n,n)-probability}
&=
\left\{
\begin{array}{cllcc}
 p, & \mbox{if}& (s,v_1,v_2,\ldots,v_n) = (0,0,\ldots,0), 
 \medskip \\
\displaystyle \frac{1-p}{t^k-1}, &\mbox{if}& (s,v_1,v_2,\ldots,v_n) \neq  (0,0,\ldots,0)\\
&& \mbox{and}~(s,v_1,v_2,\ldots,v_n)\in{\sf DT}_{(k,n)},
 \medskip \\
 0, & \mbox{if}&(s,v_1,v_2,\ldots,v_n)\not\in{\sf DT}_{(k,n)}. 
\end{array}
\right.
\end{align}
where $p \ge 1/t^k$. 

\begin{rem}
Let $\varphi: (\mathbb{F}_t)^k \to (\mathbb{F}_t)^{n+1}$ be the mapping defined by $\varphi ((s,r
_1,r_2,\ldots,r_{k-1}))=(s,v_1,v_2,\ldots,v_n)$ where $s$, $r_1,r_2,\ldots,r_{k-1}$ and $v_1,v_2,\ldots,v_n$
 are specified in \eqref{eq:relation}. Then, it is seen that $\varphi$ is injective due to the Lagrange interpolation, $\varphi((0,0,\ldots,0))=(0,0,\ldots,0)$, and $\mbox{Im}\, \varphi={\sf DT}_{(k,n)}$. Hence, we have $|{\sf DT}_{(k,n)}|= t^k$. From this fact, it is easy to see that \eqref{eq:(n,n)-probability} actually forms a probability distribution. 
\end{rem}
\begin{thm}\label{thm:nontrivaial_ideal_SSS} In the above construction $\varPi_2$, it holds for arbitrary $\ca{F} \subset [n]$ satisfying $|\ca{F}| = k-1$ that  
\begin{align} 
\nonumber
R_\infty(S|V_{\ca{F}}) &= R_\infty(S) \\
\nonumber
&= R_\infty(V_1) = \cdots = R_\infty(V_n)\\
& = -\log\frac{pt^k+(1-p)t^{k-1}-1}{t^k-1},
\end{align}
which means that, for every integers $k$ and $n$, and for arbitrary finite field of secret information, there exists a {\em non-uniform} distribution $P_S(\cdot) \in {\mathscr P}(\ca{S})$ to realize an ideal  non-perfect $\acr{\cdot}{\cdot}$-secure SSS. 
\end{thm}

{\em Proof.} 
See Appendix \ref{app:nontrivaial_ideal_SSS}. \QED

\begin{rem}
In SSSs satisfying $H(\cdot|\cdot)$-security, the ideal one exists only in the cases that the distribution of a secret is uniform or deterministic, and hence, $H(S)=H(V_i)=\log |\ca{S}|$ or $H(S)=H(V_i)=0$ is allowed in ideal  $H(\cdot|\cdot)$-secure SSSs. On the other hand, it is interesting to note that, for every $0 \le R \le \log|\ca{S}|$, there exists an ideal non-perfect SSS satisfying $\acr{\cdot}{\cdot}$-security that attains $R=R_\infty(S)$. 
\end{rem}

\section*{Acknowledgments} 
The authors are grateful to Prof.\ Hirosuke Yamamoto with The University of Tokyo for bringing their attention to \cite{M-it03}. 
The work of  Mitsugu Iwamoto is partially supported by JSPS KAKENHI Grant No.\ 23760330 and 26420345. 


\begin{thebibliography}{10}

\bibitem{S-cacm79}
A.~Shamir, ``How to share a secret,'' {\em Communications of the ACM}, vol.~22,
  no.~11, pp.~612--613, 1979.

\bibitem{B-afips79}
G.~R. Blakley, ``Safeguarding cryptographic keys,'' {\em AFIPS 1979 National
  Computer Conference}, vol.~48, pp.~313--317, 1979.

\bibitem{S-BTJ49}
C.~E. Shannon, ``Communication theory of secrecy systems,'' {\em Bell Tech.\
  J.}, vol.~28, pp.~656--715, Oct. 1949.

\bibitem{M-it03}
N.~Merhav, ``A large-deviations notions of perfect secrecy,'' {\em IEEE Trans.\
  Information Theory}, vol.~30, no.~2, pp.~506--508, 2003.

\bibitem{AS-icits12}
M.~Alimomeni and R.~Safavi-Naini, ``Guessing secrecy,'' {\em Proc.\ of the 6th
  International Conference on Information Theoretic Security (ICITS 2012),
  LNCS7412, Springer-Verlag}, pp.~1--13, August 2012.

\bibitem{D-icits12}
Y.~Dodis, ``Shannon impossibility, revisited,'' {\em Proc.\ of the 6th
  International Conference on Information Theoretic Security (ICITS 2012), {\rm
  LNCS7412, Springer-Verlag}}, pp.~100--110, August 2012.
\newblock IACR Cryptology ePrint Archive (preliminary short version): {\tt
  http://eprint.iacr.org/2012/053}.

\bibitem{Vernam-jaiee26}
G.~S. Vernam, ``Cipher printing telegraph systems for secret wire and radio
  telegraphic communications,'' {\em J.\ of American Institute for Electrical
  Engineering}, vol.~45, pp.~109--115, 1926.

\bibitem{IS-icits13}
M.~Iwamoto and J.~Shikata, ``Information theoretic security for encryption
  based on conditional {R}\'enyi entropies,'' {\em Proc.\ of International
  Conference on Information Theoretic Security (ICITS), {\rm LNCS8317,
  Springer-Verlag}}, pp.~103--121, 2013.
\newblock Full version is available from {\tt http://eprint.iacr.org/2013/440}.

\bibitem{BM-crypto85}
G.~R. Blakley and C.~Meadows, ``Security of ramp schemes,'' {\em Advances in
  Cryptology--CRYPTO'84, { \rm LNCS 196, Springer-Verlag}}, pp.~242--269, 1985.

\bibitem{HY-ieice}
H.~Yamamoto, ``On secret sharing systems using $(k,{L},n)$ threshold scheme,''
  {\em IECE. Trans.}, vol.~J68--A, no.~9, pp.~945--952, 1985.
\newblock (in Japanese). English translation: Electronics and Communications in
  Japan, Part I, vol. 69, no. 9, pp. 46--54, Scripta Technica, Inc., 1986.

\bibitem{KOSOT-ecrypt93}
K.~Kurosawa, K.~Okada, K.~Sakano, W.~Ogata, and T.~Tsujii, ``Nonperfect secret
  sharing schemes and matroids,'' {\em Advances in Cryptology--EUROCRYPT'93, {
  \rm LNCS 765, Springer-Verlag}}, pp.~126--141, 1993.

\bibitem{ISN-jc93}
M.~Itoh, A.~Saito, and T.~Nishizeki, ``Multiple assignment scheme for sharing
  secret,'' {\em J.\ of Cryptology}, vol.~6, pp.~15--20, 1993.
\newblock Preliminary version: {\it IEEE Globecom'87}, pp.99--102.

\bibitem{BSV-ipl98}
C.~Blundo, A.~D. Santis, and U.~Vaccaro, ``On secret sharing schemes,'' {\em
  Information Processing Letters}, no.~65, pp.~25--32, 1998.

\bibitem{R-BSMS61}
A.~R\'enyi, ``On measures of information and entropy,'' {\em Proc.\ of the 4th
  Berkeley Symposium on Mathematics, Statistics and Probability 1960},
  pp.~547--561, 1961.

\bibitem{A-CMSJB75}
S.~Arimoto, ``Information measures and capacity of order $\alpha$ for discrete
  memoryless channels,'' {\em Colloquia Mathematica Societatis J\'anos Bolyai,
  16. Topics in Information Theory}, pp.~41--52, 1975.

\bibitem{TAML-it12}
A.~Teixeira, A.~Matos, and L.~Antunes, ``Conditional {R}\'enyi entropies,''
  {\em IEEE Trans.\ Information Theory}, vol.~58, pp.~4273--4277, July 2012.

\bibitem{DKRS-08}
Y.~Dodis, J.~Katz, L.~Reyzin, and A.~Smith, ``Fuzzy extractors: How to generate
  strong keys from biometrics and other noisy data,'' {\em SIAM Journal on
  Computing}, vol.~38, no.~1, pp.~97--139, 2008.

\bibitem{A-it96}
E.~Arikan, ``An inequality on guessing and its application to sequential
  decoding,'' {\em IEEE Trans.\ Information Theory}, vol.~42, no.~1,
  pp.~99--105, 1996.

\bibitem{KGH-it83}
E.~D. Karnin, J.~W. Greene, and M.~E. Hellman, ``On secret sharing systems,''
  {\em IEEE Trans. Inform. Theory}, vol.~29, no.~1, pp.~35--41, 1983.

\bibitem{CSGV-jc93}
R.~M. Capocelli, A.~D. Santis, L.~Gargano, and U.~Vaccaro, ``On the size of
  shares for secret sharing schemes,'' {\em Journal of Cryptology}, vol.~6,
  pp.~157--167, 1993.

\bibitem{H-it11}
M.~Hayashi, ``Exponential decreasing rate of leaked information in universal
  random privacy amplification,'' {\em IEEE Trans.\ Information Theory},
  vol.~57, no.~6, pp.~3989--4001, 2011.

\bibitem{BL-ecrypt88}
J.~Benaloh and J.~Leichter, ``Generalized secret sharing and monotone
  functions,'' {\em Advances in Cryptology--CRYPTO'88, { \rm LNCS 403,
  Springer-Verlag}}, pp.~27--35, 1990.

\bibitem{Stinson-05}
D.~R. Stinson, {\em CRYPTOGRAPHY Theory and Practice}.
\newblock CRC Press, third~ed., 2005.

\end{thebibliography}

\appendix

\subsection{Proof of Theorem \ref{thm:n-out-of-n-SSS}}\label{sec:n-out-of-n-SSS}

First, we show \eqref{eq:ideality} and \eqref{eq:non-ideality}. It is easy to see that $R_\infty(S)\allowbreak  = R_\infty(V_1)=\cdots=R_\infty(V_{n-1})= -\log p$. Noticing that $V_n=S\oplus V_1\oplus \cdots\oplus V_{n-1}$, it holds that $S\oplus V_1 = V_n$, and hence, we have $P_{V_n}(0)=p^2+q^2$ where $q:=1-p$. Since $p \ge q$, we obtain  $R_\infty (V_n) = -\log(p^2+q^2)$. 

In the remaining of this proof, we check \eqref{eq:security_pi1}. 
Since $R_\infty(X|Y)$ satisfies Corollary \ref{prop:basic-properties} (A), it is sufficient to show 
$R_\infty(S|V_{\ca{F}}) = R_\infty(S)$ only in the case of $|\ca{F}|=n-1$. 

 Consider the case where $n \not\in \ca{F}$. In  this case, it is easy to see that 
$$
P_{S|V_1V_2\cdots V_{n-1}}(s|v_1,v_2,\ldots,v_{n-1})=P_S(s)
$$
holds since $S$ and $V_1,V_2,\ldots,V_{n-1}$ are independent. Hence, $R_\infty(S|V_{\ca{F}}) = R_\infty(S)$ obviously holds in this case.

Next, we consider the case of $n \in \ca{F}$. From the symmetricity, it is sufficient to consider the case where $\ca{F}=\{2,3,\ldots,n\}$. For simplicity of notation, define\footnote{In the case of $n=2$, we set $\v{v}=\emptyset$ and $\v{V}=\emptyset$.}    $\v{v}:=v_{\ca{F}\backslash\{n\}}$ and  $\v{V}:=V_{\ca{F}\backslash\{n\}}$. 
Let $\sigma: \{0,1\}^{n-2} \rightarrow \{0,1\}$ be the mapping that computes exclusive OR of all inputs. Due to the construction and the independency among $S$ and $V_1,V_2,\ldots,V_{n-1}$, the probability $P_{V_\ca{F}}(v_{\ca{F}})=P_{\sv{V}V_n}(\v{v},v_n)$ can be calculated in the following cases: 
\begin{description}
\item[Case 1):] ~~~$\sigma(\v{v})\oplus v_n=1$, i.e., $(\sigma(\v{v}),v_n)=(0,1)$ or $(\sigma(\v{v}),v_n)=(1,0)$:
\begin{align}
\nonumber
P_{\sv{V}V_n}(\v{v},v_n)&=P_{SV_1\sv{V}}(1,0,\v{v})+P_{SV\sv{V}}(0,1,\v{v})\\
&=2pq P_{\sv{V}}(\v{v})
\end{align}
\item[Case 2):] ~ $\sigma(\v{v})\oplus v_n=0$, i.e., $(\sigma(\v{v}),v_n)=(0,0)$ or $(\sigma(\v{v}),v_n)=(1,1)$:
\begin{align}
\nonumber
P_{\sv{V}V_n}(\v{v},v_n)&=P_{SV_1V}(0,0,\v{v})+P_{SV_1V}(1,1,\v{v})\\
&=(p^2+q^2)P_{\sv{V}}(\v{v})
\end{align}
\end{description}
Furthermore, note that the following relation: 
\begin{align}
\nonumber
P_{S|\sv{V}V_n}(s|\v{v},v_n)&=P_{SV_1|\sv{V}V_n}(s,1|\v{v},v_n)\\
&~~~~+P_{SV_1|\sv{V}V_n}(s,0|\v{v},v_n). 
\label{eq:prob_SSS}
\end{align}

Now, consider Case 1). In this case it is easy to see that 
\begin{align}
P_{SV_1|\sv{V}V_n}(0,1|\v{v},v_n)&=P_{SV_1|\sv{V}V_n}(1,0|\v{v},v_n)=\frac{1}{2},
\end{align}
and
\begin{align}P_{SV_1|\sv{V}V_n}(0,0|\v{v},v_n)=P_{SV_1|\sv{V}V_n}(1,1|\v{v},v_n)=0. 
\end{align}
Hence, \eqref{eq:prob_SSS} becomes 
$P_{S|\sv{V}V_n}(s|\v{v},v_n)=1/2$, which leads to
\begin{align}
\label{eq:case1}
P_{\sv{V}V_n}(\v{v},v_n)\max_s P_{S|\sv{V}V_n}(s|\v{v},v_n)=pq \cdot P_{\sv{V}}(\v{v}). 
\end{align}

Next, consider  Case 2). In this case, it is easy to see that 
$$
P_{SV_1|\sv{V}V_n}(0,1|\v{v},v_n)=P_{SV_1|\sv{V}}(1,0|\v{v},v_n)=0
$$
and hence, \eqref{eq:prob_SSS} becomes 
\begin{align}
\nonumber
&\max_s P_{S|\sv{V}V_n}(s|\v{v},v_n)\\
\label{eq:prob_SSS2}
&=\max\left\{P_{SV_1|\sv{V}V_n}(0,0|\v{v},v_n),P_{SV_1|\sv{V}V_n}(1,1|\v{v},v_n)\right\}.
\end{align}
 
Here, $P_{SV_1|\sv{V}V_n}(0,0|\v{v},v_n)$ can be calculated as follows:
\begin{align}
\nonumber
P_{SV_1|\sv{V}V_n}(0,0|\v{v},v_n) &= \frac{P_{SV_1\sv{V}V_n}(0,0,\v{v},v_n)}{P_{\sv{V}}(\v{v},v_n)}\\
\nonumber
&= \frac{P_{SV_1}(0,0)P_{\sv{V}}(\v{v})}{P_{\sv{V}}(\v{v},v_n)}\\
&= \frac{p^2}{p^2+q^2}
\end{align}
Similarly, we have $P_{SV_1|\sv{V}V_n}(1,1|\v{v},v_n) =  q^2/(p^2+q^2)$. 
Hence, because of $p \ge q$, \eqref{eq:prob_SSS2} becomes 
\begin{align}
\label{eq:case2}
P_{\sv{V}V_n}(\v{v},v_n)\max_s P_{S|\sv{V}}(s|\v{v}) = p^2 \cdot P_{\sv{V}}(\v{v}). 
\end{align}
 
Summarizing \eqref{eq:case1} and \eqref{eq:case2}, we have 
\begin{align}
\nonumber
&\sum_{\sv{v},v_n} P_{\sv{V}V_n}(\v{v},v_n)\max_s P_{S|\sv{V}V_n}(s|\v{v},v_n)\\
\nonumber
&=
 \sum_{\sv{v}:\sigma(\sv{v})=0\atop v_n=1} 
 \sum_{\sv{v}:\sigma(\sv{v})=1\atop v_n=0} pq P_{\sv{V}}(\v{v})+
 \sum_{\sv{v}:\sigma(\sv{v})=0\atop v_n=0} 
 \sum_{\sv{v}:\sigma(\sv{v})=1\atop v_n=1} p^2 P_{\sv{V}}(\v{v})\\
&=
 \sum_{\sv{v}:\sigma(\sv{v})=0,1} (pq + p^2)P_{\sv{V}}(\v{v})=p. 
\end{align}

Hence, we obtain $R_\infty(S|V_{\ca{F}})=R_\infty(S)=-\log p$, which completes the proof. \hfill \QED

\subsection{Proof of Theorem \ref{thm:nontrivaial_ideal_SSS}}\label{app:nontrivaial_ideal_SSS}

\noindent
Since the joint probability of $(S,V_1,V_2,\ldots,V_n)$ is $\frac{1-p}{t^k-1}$ except the case where $S,V_1,V_2,\ldots,V_n$ are all $0$, it is easy to see that
\begin{align}
\nonumber
P_S(0)&=P_{V_1}(0)=P_{V_2}(0) = \cdots =P_{V_n}(0) \\
\nonumber
&= p+(t^{k-1}-1)\frac{1-p}{t^k-1}\\
\nonumber
&=p-\frac{1-p}{t^k-1}+t^{k-1}\frac{1-p}{t^k-1}\\
\label{eq:P_S(0)}
&=\frac{pt^k+(1-p)t^{k-1}-1}{t^k-1}\\
\nonumber
P_S(z)&=P_{V_1}(z)=P_{V_2}(z) = \cdots =P_{V_n}(z) \\
\label{eq:P_S(1)}
&= t^{k-1}\frac{1-p}{t^k-1},~\mbox{ for }~z\in[t-1]
\end{align}

Here we note that $p \ge 1/t^k$ implies that $p \ge \frac{1-p}{t^k-1}$. 
Comparing with \eqref{eq:P_S(0)} and \eqref{eq:P_S(1)} taking $p \ge \frac{1-p}{t^k-1}$ into account, it is easy to see that $P_S(0) \ge P_S(z)$ as well as  $P_{V_i}(0) \ge P_{V_i}(z)$, $i\in[n]$, for arbitrary $z\in\mathbb{F}_t^+:=\mathbb{F}_t\backslash\{0\}$. Hence, we have
\begin{align} 
\nonumber
R_\infty(S) &= R_\infty(V_1) = \cdots = R_\infty(V_n)\\
& = -\log\frac{pt^k+(1-p)t^{k-1}-1}{t^k-1}
\end{align}

Now, we calculate $H(S|V_{\ca{F}})$ where we assume that $V_{\ca{F}}:= \{V_1,V_2,\ldots,V_{k-1}\}$ without loss of generality. 

First, we consider the case where $(v_1,v_2,\ldots,v_{k-1})=(0,0,\ldots,0)$ is the condition. In the case, we have
\begin{align}
\nonumber
P_{V_1V_2\cdots V_{k-1}}(0,0,\ldots,0)&=p+(t-1)\frac{1-p}{t^k-1}
\end{align}
and we denote this probability by $R(p,t,k)$ for simplicity. 
Hence, we have 
\begin{align}
\nonumber
P_{S|V_1V_2\cdots V_{n-1}}&(s|0,0,\ldots,0)\\
& = 
\left\{
\begin{array}{rcccc}
p/R(p,t,k),&\mbox{if}& s=0\\
\frac{1-p}{t^k-1}/R(p,t,k),&\mbox{if}& s=1\\
\end{array}
\right.
\end{align}
which results in 
$\max_{s\in \mathbb{F}_t} P_{S|V_1V_2\cdots V_{t-1}}(s|0,0,\ldots,0) = p/R(p,t,k)$ 
since we assume that $p \ge 1/t^k$. 

On the other hand, consider the case of $(v_1,v_2,\ldots,v_{k-1})\neq (0,0,\ldots,0)$. In this case, we have 
\begin{align}
P_{V_1V_2\cdots V_{k-1}}(v_1,v_2,\ldots,v_{k-1})=t\frac{1-p}{t^k-1}.
\end{align}
and hence, it holds that 
\begin{align}
P_{S|V_1V_2\cdots V_{n-1}}(s|v_1,v_2,\ldots,v_{k-1}) = \frac{1}{t}
\end{align}
and hence, we obtain  
$\max_s P_{S|V_1V_2\cdots V_{n-1}}(s|v_1,v_2,\ldots,v_{n-1}) \allowbreak = 1/t$. 

Summarizing, we have
\begin{align}
\nonumber
H(S&|V_1V_2\cdots V_{n-1}) \\
\nonumber
&= -\log\left\{
p\Big/R(p,t,k)\times R(p,t,k) \right.\\
\nonumber
&~~~~~~~~~~~~\left.
+\frac{1}{t} \times t\frac{1-p}{t^k-1} \times (t^{k-1}-1)
\right\}\\
\nonumber
&= -\log\left\{p+(t^{k-1}-1)\frac{1-p}{t^k-1}\right\}\\
&=-\log\frac{pt^k+(1-p)t^{k-1}-1}{t^k-1}.
\end{align}

\end{document}